\documentclass{jltp}
\usepackage{graphicx} % uncomment this line to include the graphicx package

\title{Why Study Noise due to Two Level Systems: A Suggestion for
Experimentalists}

\author{Clare C. Yu}

\address{Department of Physics and Astronomy, University of California,
Irvine, California 92697}

\runninghead{C. C. Yu}{Noise due to Two Level Systems}

\begin{document}

\maketitle

\begin{abstract}
Noise is often considered to be a nuisance. Here we argue that it can
be a useful probe of fluctuating two level systems in glasses.
It can be used to: (1) shed light on whether the fluctuations are correlated or
independent events; (2) determine if there is a low temperature
glass or phase transition among interacting two level systems, and if
the hierarchical or droplet model can be used to describe the glassy
phase; and (3) find the lower bound of the two level system relaxation
rate without going to ultralow temperatures. Finally we point out
that understanding noise due to two level systems is important for
technological applications such as quantum qubits that use Josephson
junctions.

\end{abstract}

\section{INTRODUCTION}
It is a great pleasure to contribute a paper in honor of Professor 
Sigfried Hunklinger.
I first met Professor Hunklinger in 1986 when I was a postdoc
at the University of Illinois in Urbana. Andy Anderson and Jim Wolfe had
organized a Phonon Scattering Conference, and due to a last minute
cancellation, I was scheduled to give the first contributed talk.
Andy made it quite clear that I should do a good job so that the conference
would start off on the right foot, so I was a little nervous. Sigfried
was the chair of the session. Before the session started,
he came up to me, smiled, and introduced himself, saying that he wanted
to make sure to pronounce my name correctly. The talk went fine. 
 
The next Phonon Scattering Conference was in Heidelberg in 1989
and Sigfried was one of the organizers. In spite of being busy
with running the conference, he managed to stop by my poster
where I was presenting some work I had done with Tony Leggett
proposing that interactions between two level systems were important. 
This was not the prevailing view at the time.
As he left, he smiled, waved the paper I had given him, and said, 
``I believe it.''
I would like to thank Professor Hunklinger for his encouragement
through the years. His enthusiasm has helped to make
the field of glasses at low temperatures a fun field.

I have found Professor Hunklinger's
papers to be a source of learning and inspiration through the years.
I still use his review article with Raychaudhuri\cite{hunk} 
as a basic reference to two level systems. His recent work
on the effect of magnetic fields provoked a great deal of interest and
thoughtful discussion in the field. 
The following is based on a talk I gave in June 2003 in Dresden at the 
International Workshop on Collective Phenomena in the Low Temperature
Physics of Glasses that Professor Hunklinger helped to organize.

To an experimentalist, noise is a nuisance at best and a serious problem
hindering measurements at worst. However noise comes from the fluctuations
of microscopic entities and it can act as a probe of what is happening
physically at the microscopic scale. Utilizing noise has been done in 
electronic systems where conductivity can be 
measured\cite{Kogan96,Dutta81,Weissman88}. In this paper 
I would like to argue
that this should also be done in measurements of the dielectric function
in glasses at low temperatures. Let us assume that fluctuations
in the dielectric function are due to fluctuations in two level
systems making transitions from one state to another. Measuring the
noise could help to determine if the fluctuations are correlated
or independent. It could also be used to find the lower bound on
the two level system (TLS) relaxation rate due to phonon emission, or 
equivalently, the lower bound $\Delta_o^{min}$ on the TLS tunneling
matrix element, without going to ultralow temperatures that are of order
$\Delta_o^{min}$. There has been speculation that interacting
two level systems should undergo a phase transition as the
temperature is lowered. Professor Hunklinger was involved in
finding experimental evidence suggesting this\cite{Strehlow98}.
Studying the noise may help to determine if there is a glass transition or
phase transition of interacting two level systems. If such a
transition does occur, the second spectrum of the noise can
be used to determine if this phase can be described by the
hierarchical or droplet model\cite{Weissman92,Weissman93}.
(We will explain what a second spectrum is later.) Finally
noise due to two level systems could be technologically important.
For example, two level systems in the oxide layer that acts
as a tunnel barrier in a Josephson junction can have deleterious
effects on quantum qubits that use Josephson 
junctions\cite{Martinis04}.

\subsection{Introduction to Noise}
Let us set up our notation and define what we mean by noise.
Let $p(t)$ be a quantity that fluctuates in time.
Let $\delta p(t)$ be the deviation from its average value 
of some quantity $p$ at time $t$.
If the processes producing the fluctuations are stationary in time, i.e.,
translationally invariant in time, then
the autocorrelation function of the fluctuations
$\langle\delta p(t_2)\delta p(t_1)\rangle$ will be a function 
$\psi(t_2-t_1)$
of the time difference. In this case the Wiener--Khintchine theorem
can be used to relate the noise spectral density $S_p(\omega)$ to the
Fourier transform $\psi(\omega)$ of the autocorrelation 
function\cite{Kogan96}: $S_{p}(\omega)=2\psi_{p}(\omega)$
where $\omega$ is the angular frequency. In practice $S_p(\omega)$
typically is calculated by multiplying the time series $\delta p(t)$
by a windowing or envelope function so that the time series goes
smoothly to zero, Fourier transforming the result, taking the modulus
squared, and multiplying by two to obtain the noise
power\cite{Press92}.

1/f noise, which is ubiquitous and
dominates at low frequencies, corresponds to $S_p(\omega)\sim 1/\omega$.
A simple way to obtain 1/f noise was given by Dutta and Horn\cite{Dutta81}.
We can use
the relaxation time approximation to write the equation of motion for
$\delta p$:
\begin{equation}
\frac{d\delta p}{dt}=-\frac{\delta p}{\tau}
\end{equation}
where $\tau^{-1}$ is the relaxation rate. The autocorrelation function
$\psi_{p}(t)$ is given by
\begin{equation}
\psi_{p}(t)=\langle\delta p(t)\delta p(t=0)\rangle
\end{equation}
The Fourier transform is a sum of Lorentzians:
\begin{equation}
\psi_{p}(\omega)=A\int^{D}_{-D}\frac{\tau(\varepsilon,T)}
{1+\omega^2\tau^{2}(\varepsilon,T)}g(\varepsilon,T)d\varepsilon
\label{eq:lorentzians}
\end{equation}
where we assume that the relaxation time $\tau$ is a function of
energy $\varepsilon$. The range of energies has a bandwidth of $2D$.
$A$ is an overall scale factor. 
If we use $g(\varepsilon,T)=g_o$ for the density of states and
$\tau=\tau_o\exp\left(\varepsilon/k_BT\right)$ where $g_o$ and
$\tau_o$ are constants,
$T$ is temperature and $k_B$ is Boltzmann's constant, the noise spectral
density is given by
\begin{equation}
S(\omega)=2\psi_{p}(\omega)\sim\frac{1}{\omega}
\end{equation}
Thus we obtain 1/f noise from independent fluctuators with a distribution
of relaxation times.

\subsection{Two Level System Model}
The standard model of noninteracting two level 
systems\cite{hunk,phillips} was introduced
by Anderson, Halperin, and Varma\cite{ahv}, and independently
by W. A. Phillips\cite{tls} in 1972. Let us briefly review this model
in order to set up the notation that we will use.
The standard Hamiltonian for a two level system is 
\begin{equation}
H=\frac{1}{2} \left( \begin{array}{cc} 
\Delta & \Delta_o \\ 
-\Delta_o & -\Delta 
\end{array} \right) 
\end{equation}

Here we are using the left well -- right well basis.
$\Delta$ is the asymmetry energy, i.e., $\Delta$ is the
energy difference between the right well and the left well.
$\Delta_o$ is the tunneling matrix element and is given by
\begin{equation}
\Delta_o = \hbar\omega_o e^{-\lambda}
\end{equation}
where $\omega_o$ is the attempt frequency and is typically
of order the Debye frequency. The exponent $\lambda$ is given
by
\begin{equation}
\lambda=\sqrt{\frac{2mV}{\hbar^2}}d
\end{equation}
where $m$ is the mass of the tunneling entity,
$V$ is the barrier height and $d$ is the distance between wells.
In the standard model of noninteracting two level systems $\Delta$ and 
$\lambda$ have flat distributions:
\begin{equation}
P(\Delta,\lambda)d\Delta d\lambda = P_o d\Delta d\lambda
\end{equation}
where $P_o$ is a constant. The distribution $P(\Delta,\Delta_o)$ is
given by
\begin{equation}
P(\Delta,\Delta_o)d\Delta d\Delta_o=\frac{P_o}{\Delta_o}d\Delta d\Delta_o
\end{equation}
We can diagonalize the Hamiltonian to get the energy eigenvalues
that are given by $\pm \varepsilon/2$ where 
\begin{equation}
\varepsilon=\sqrt{\Delta^2 + \Delta_o^2}
\end{equation}
The relaxation rate for an excited two level system to emit a phonon
and return to its ground state is given by\cite{phillips,hunk}
\begin{equation}
\tau^{-1}_{1}=\frac{\gamma^2}{\rho}\left[\frac{1}{c^{5}_{\ell}}+
\frac{2}{c^{5}_{t}}\right]\frac{\varepsilon^{3}}{2\pi\hbar^{4}}
\left[\frac{\Delta_{o}}{\varepsilon}\right]^{2}\coth
\left[\frac{\beta\varepsilon}{2}\right]
\label{eq:tau}
\end{equation}
where $\gamma$ is the deformation potential, $\rho$
is the mass density, $c_{\ell}$ is the longitudinal speed of
sound, $c_{t}$ is the transverse speed of sound, and $\beta=1/k_BT$.

\section{1/f NOISE DUE TO TWO LEVEL SYSTEMS}
\subsection{Noninteracting Two Level Systems}
\begin{figure}
\centerline{\includegraphics[height=3.5in]{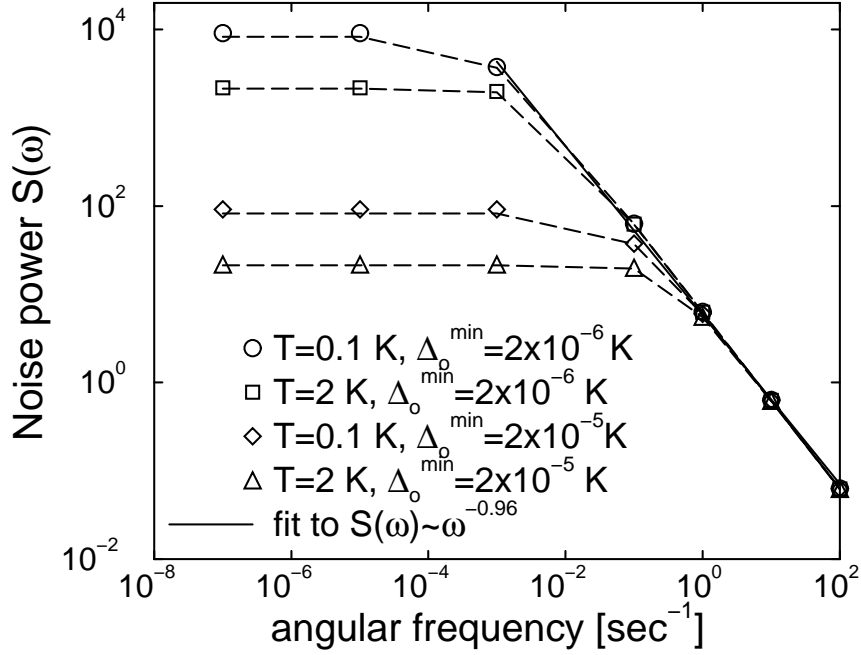}}
\caption{Log-log plot of noise power vs. frequency $\omega$ from
two level systems for T = 0.1 K and T=2 K, and
for $\Delta_{o}^{min}=2\times 10^{-5}$ K and $2\times 10^{-6}$ K.
The open symbols represent the spectral density of the noise
due to noninteracting two level systems relaxing via phonons. 
$S(\omega)$ is
calculated using Eqs.~(\ref{eq:tau}) and (\ref{eq:noninteractingNoise}). 
The dashed lines show the 
noise power due to two level systems
relaxing via phonons and via interactions with other two
level systems. The dashed lines are calculated using 
Eqs.~(\ref{eq:noninteractingNoise}) and (\ref{eq:interactingTLS}).
The solid line is a fit to $S(\omega)\sim\omega^{-0.96}$ and shows
that two level systems produce 1/f noise at high frequencies.
We have used the values for SiO$_{2}$:
$\gamma=$1 eV, $\rho=$ 2200 kg/m$^{3}$, $c_{\ell}$=5800
m/s, $c_{t}$=3800 m/s, 
$\Delta_{o}^{max}=4$ K, $\Delta^{max}$=4 K. We set the
overall scale factor $AP_o=1$. The values of
$\Delta_{o}^{min}$ are indicated in the figure. For the case of
interacting two level systems, we set $T_C=55$ mK which is the value
for Mylar\cite{Ludwig03}.} 
\label{fig:noninteractingNoise}
\end{figure}

Many two level systems have elecric dipole moments associated with
them. Fluctuating two level systems produce noise in the dielectric
polarization. To find the noise power produced by noninteracting two level
systems, we can plug the relaxation time from Eq.~(\ref{eq:tau})
into Eq.~(\ref{eq:lorentzians}):
\begin{equation}
S(\omega)=AP_{o}\int^{\Delta^{max}}_{-\Delta^{max}}d\Delta
\int^{\Delta_{o}^{max}}_{\Delta_{o}^{min}}
\frac{d\Delta_{o}}{\Delta_{o}}\frac{\tau\left(\Delta,\Delta_{o},T\right)}
{1+\omega^{2}\tau^{2}\left(\Delta,\Delta_{o},T\right)}
\label{eq:noninteractingNoise}
\end{equation}
where $\Delta^{max}$ is the maximum value of the asymmetry energy, and
$\Delta_{o}^{min}$ and $\Delta_{o}^{max}$ are the minimum and maximum
values of the tunneling matrix element, respectively. The result of 
evaluating Eq.~(\ref{eq:noninteractingNoise})
is shown in Figure \ref{fig:noninteractingNoise}. At high frequencies, 
the noise power is 1/f because
it can be fit quite well with $S(\omega)\sim \omega^{-0.96}$.

\begin{figure}
\centerline{\includegraphics[height=2.5in]{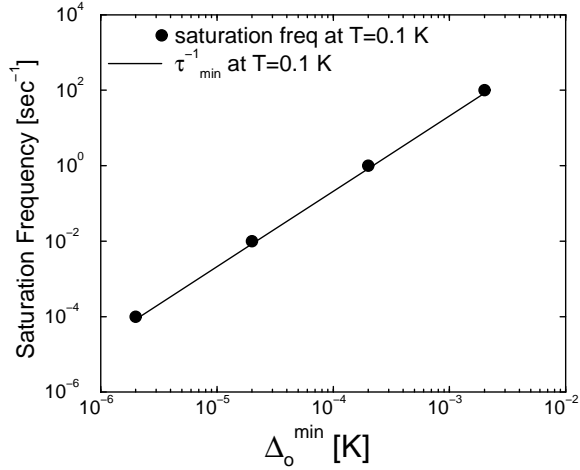}}
\caption{Log-log plot of saturation frequency vs. 
$\Delta_{o}^{min}$ at T=0.1 K for noninteracting two level systems. 
$\tau^{-1}_{min}$ is calculated from Eq.~(\ref{eq:tau}) with
$\varepsilon=\Delta_{o}^{min}$. The values of the parameters
are the same as in Figure \ref{fig:noninteractingNoise}.}
\label{fig:saturation}
\end{figure}

The noise saturates at low frequencies. The saturation frequency
where the plot rolls over is given by the inverse of the 
longest relaxation time $\tau^{-1}_{min}$, and hence depends 
on $\Delta_{o}^{min}$ and the temperature.  
In Figure \ref{fig:saturation} we show a plot of the saturation
frequency as a function of $\Delta_{o}^{min}$. 
Although we do not have a precise definition of the saturation 
frequency, we identify it to be the frequency below which the noise
power saturates.  Figure \ref{fig:saturation} shows that the saturation
frequency increases as $\Delta_{o}^{min}$ increases. 
Figure \ref{fig:saturation} also shows the minimum relaxation rate
$\tau^{-1}_{min}$ calculated using Eq.~(\ref{eq:tau}) with
$\varepsilon=\Delta_{o}^{min}$. We see that saturation frequency
matches $\tau^{-1}_{min}$ as expected. 

When averaging over the distribution of parameters for two level systems, 
$\Delta_{o}^{min}$ is needed as a lower limit to prevent the $\Delta_{o}$ 
integral from diverging. In addition interactions between two level
systems should produce a hole in the distribution of energy splittings
\cite{herve}
which could be interpreted as evidence of 
the existence of $\Delta_{o}^{min}$. There has been a long
standing question of whether $\Delta_{o}^{min}$ exists and if so,
what a reasonable value of it is. Finding whether or not 1/f noise
saturates at low frequencies would provide a way to determine
$\Delta_{o}^{min}$ without going to ultralow temperatures of order
$\Delta_{o}^{min}$. Of course a very small value of $\Delta_{o}^{min}$
would correspond to extremely low frequencies that may not be
experimentally accessible. Typical values are shown in 
Figure \ref{fig:saturation}.

\subsection{Interacting Two Level Systems}
Two level systems can interact with one another via strain fields and via
dipole--dipole interactions if they have electric dipole 
moments\cite{ccyajl,herve,Esquinazi98}.
So we can ask about the noise spectrum produced by interacting two
level systems. There are several ways to approach this. One way
would be to allow an excited two level system to relax to its
ground state by emitting a phonon or by transferring its excitation
energy to other two level systems. In this case the relaxation rate
is given by
\begin{equation}
\tau^{-1}_{1}=\tau^{-1}_{1,phonons}+
B_o\left(\frac{\Delta_{o}}{\varepsilon}\right)^{2} k_BT
\label{eq:interactingTLS}
\end{equation}
where the second term represents the relaxation rate via energy transfer
to other two level systems\cite{Burin95}. $B_o$ is a constant given by  
\begin{equation}
B_o=\frac{\gamma^2}{\rho}\left[\frac{1}{c^{5}_{\ell}}+
\frac{2}{c^{5}_{t}}\right]\frac{\left(k_BT_C\right)^{2}}{2\pi\hbar^{4}}
\left.\coth\left(E/k_BT\right)\right|_{E=k_BT}
\label{eq:Bo}
\end{equation}
where we set $E=k_BT$ in Eq.~(\ref{eq:Bo}) because the two level
systems that obey $E=k_BT$ are the dominant contribution to the
dielectric function\cite{Nalbach04}. The reason for this
is given by Nalbach {\it et al.}\cite{Nalbach04}. The resonant 
dielectric response of two level systems has a thermal occupation 
factor $\tanh\left(E/k_BT\right)$. This implies that at high temperatures
($k_B T>E$) both states are equally populated and there is no net
polarization. Thus two level systems
with $E\stackrel{>}{\sim}k_BT$ dominate.
On the other hand, only two level systems with
$E\stackrel{<}{\sim} k_BT$ are accessible. So only two level systems with
$E\sim k_BT$ are relevant for changes of the dielectric constant.

$T_C$ is the
temperature where the rate of two level system
relaxation via phonon emission equals the rate of relaxation
via interactions with other two level systems, i.e., where the
two terms in Eq.~(\ref{eq:interactingTLS}) are equal. Plugging 
Eq.~(\ref{eq:interactingTLS}) into Eq.~(\ref{eq:noninteractingNoise})
by replacing $\tau\left(\Delta,\Delta_{o},T\right)$ with $\tau^{-1}_{1}$
again yields 1/f noise at high frequencies which saturates at low 
frequencies as shown in Figure \ref{fig:noninteractingNoise}. 
From the figure we see that including interactions between
two level systems has a negligible
effect on the noise. One might expect that at lower temperatures,
interactions would have a more pronounced effect, but even when
we set $T=0.02$ K, the spectral density of the noise follows almost the
same curve as for $T=0.1$ K.
However, simply substituting Eq.~(\ref{eq:interactingTLS})
into Eq.~(\ref{eq:noninteractingNoise}) is not a correct way to calculate the
noise power. The relaxation time in Eq.~(\ref{eq:lorentzians})
refers to the relaxation time of a coherent mode, not to the relaxation
time of a single two level system interacting with other two
level systems. So the correct calculation would involve
diagonalizing the system of interacting two level systems to find
the relaxation times of different normal modes to the ground state. It is
these relaxation times that one should use in Eq.~(\ref{eq:lorentzians}).

\section{SECOND SPECTRUM OF THE NOISE}

We can also look at the so--called second spectrum of the noise. To 
understand
the second spectrum, consider the following. Suppose we take a time series 
that has 500,000 points in it. We divide the time series into 50
segments, each with 10,000 points in it. Then we calculate the
first spectrum $S(\omega)$ of each segment so that we obtain
50 first spectra $S(\omega,t_{i})$ where $i=1,...,50$ and $t_{i}$
is the starting time of the $i$th time series. 
So for a given value of the frequency $\omega = \omega_{1}$, we have a time
series $S(\omega_1,t_{i})$ with 50 points. We can Fourier transform
this time series to get the second spectrum $S_2(\omega_1,\omega_2)$.
The second spectrum is the power spectrum of the fluctuations of $S(\omega)$
with time, i.e., the Fourier transform of the autocorrelation function
of the time series of $S(\omega)$\cite{Restle85,Weissman92,Weissman93}. 
What we described above is not exactly how the second 
spectrum is usually calculated.
To calculate the second spectrum, we can divide each first 
spectrum into octaves. An octave is a range of frequencies
from $\omega_L$ to $\omega_H$ where typically $\omega_H=2\omega_L$. 
We can discretize the
first spectrum by associating each octave with the total noise power in that
octave. The total power is the sum of the values $S(\omega)$
with $\omega_L < \omega < \omega_H$.
We do this for each first spectrum. For each octave this gives us a time 
series with one number from each first spectrum labeled by $t_i$. 
This time series represents the fluctuations in the noise power 
in a given octave labeled by frequency $\omega_L$, say.
We can calculate the autocorrelation function of these fluctuations,
Fourier transform it and obtain the noise power 
$S_2(\omega_L,\omega_2)$ that is the
second spectrum. Equivalently, by Parseval's theorem,
we can Fourier transform the time series of noise power fluctuations,
then take the modulus squared and multiply by two to obtain
the second spectrum. 

Rather than doing a Fourier transform for the first
spectrum $S(\omega)$, one can do a simple
wavelet transform which is known as a Haar transform\cite{Petta98}; this
is more efficient computationally. Doing a Haar transform effectively
multiplies the time series by a square wave. To do a Haar transform,
start with a time series. For the lowest frequency square wave, 
subtract the sum of the first half of the data from the sum of
the second half of the data. 
For the next higher frequency subtract the sum of the first
quarter of the data from the sum of the second quarter of the data. 
And subtract the sum of the third quarter of the data from the sum 
of the fourth quarter of the data. Keep going until you are 
subtracting the first point from the second point. Now you
have a set of Haar transforms. Square the value of each Haar point
to obtain the Haar power. For a given square wave, there are values
of the Haar power at different times. This square wave can be 
associated with a given frequency $\omega_{1}$. Fourier transform
the values of the Haar power corresponding to a given square wave.
The frequency of the Fourier transform is $\omega_2$.
Square the Fourier transform and
this gives the second spectrum $S_2(\omega_{1},\omega_2)$. 
It is customary to normalize the
second spectrum by dividing by the square of the average of the first
spectrum, i.e., the Haar power.

The second spectrum can tell us if the fluctuators are correlated or
independent\cite{Weissman88,Weissman92,Weissman93}. 
If the second spectrum is white (independent of $\omega_2$)
the fluctuators are not correlated.
If the fluctuators are correlated, then the second spectrum will be
frequency dependent. 
The frequency dependence of the second spectrum could be used to 
detect a phase transition among interacting two level systems.
If interacting two level systems undergo a glass transition or phase
transition, the second spectrum will change from being frequency
independent in the high temperature phase to being frequency
dependent in the low temperature glassy phase.

We may be able to draw a useful analogy between interacting two level
systems in glasses and spin glasses if we identify two level systems with
(Ising) spins. Measurements of the second spectrum have
been used to differentiate between the hierarchical model and the droplet
model of spin glasses because these two models assume different 
correlations between the fluctuators\cite{Weissman92,Weissman93}. 

In the droplet model, 
entire clusters or droplets of spins coherently flip and produce 
fluctuations in the magnetization \cite{Bray87,Fisher88a,Fisher88b}. 
The energy for a cluster to flip scales as $L^{\theta}$ where
$L$ is the linear size of the droplet and the power $\theta$ is
small. There are fewer large droplets than small droplets, and the 
big droplets flip less frequently than the small droplets. So large
clusters contribute to the low frequency noise and small fast clusters
contribute to the high frequency noise. In the simplest version of this
model the droplets are noninteracting. If this were the case, the
second spectrum would be frequency independent. A more sophisticated
version has interacting droplets. Large droplets are more likely
to interact than small droplets so the second spectrum will be larger
at low frequencies $\omega_1$. 

\begin{figure}
\centerline{\includegraphics[height=2.0in]{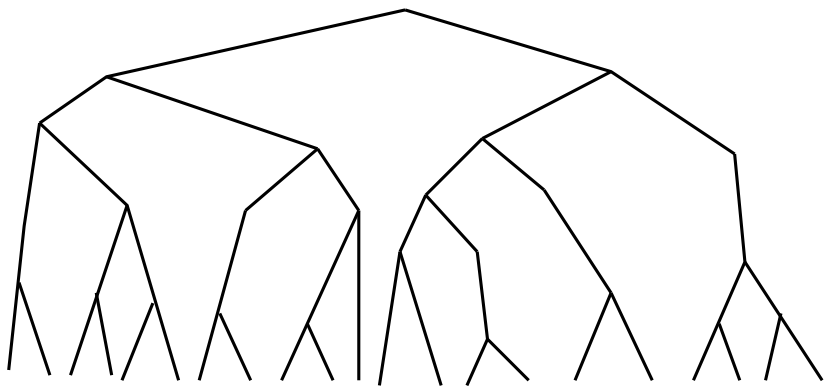}}
\caption{Bifurcating tree structure of the hierarchical model.}
\label{fig:hierarchy}
\end{figure}

In the hierarchical 
model\cite{Ogielski85,Paladin85,Schreckenberg85,Bachas86,Maritan86,Sibani87,Weissman92,Weissman93}
the states (or spin arrangements) of the spin glass lie at the endpoints
of a bifurcating hierarchical tree as shown in Figure \ref{fig:hierarchy}.
The tree structure is self--similar.
The Hamming distance $D$ between two states
is the fraction of spins that must be reoriented to convert one state into
another. It turns out that $D$ corresponds to the highest vertex
on the tree along the shortest path connecting the states. The farther
two states are, the longer the time to go between them. 

\begin{figure}
\centerline{\includegraphics[height=2.0in]{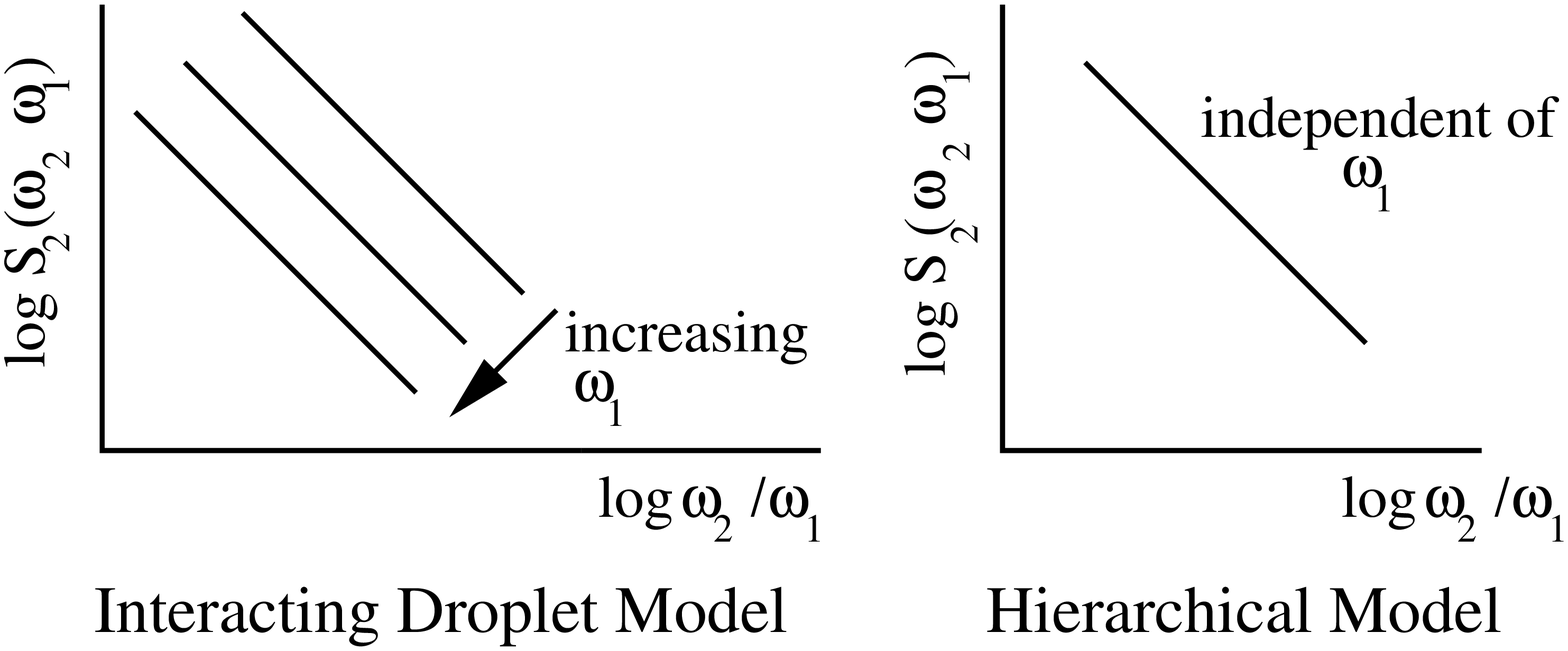}}
\caption{Sketch of the second spectrum for the interacting droplet model and
the hierarchical model as a function of the ratio $\omega_2/\omega_1$ for
different values of $\omega_1$.}
\label{fig:S2}
\end{figure}

It is often useful to plot the second spectrum $S_2(\omega_2,\omega_1)$
as a function of the ratio $\omega_2/\omega_1$. The 
hierarchical model predicts that the second spectrum should be
scale invariant and only depend on the ratio $\omega_2/\omega_1$ and not on 
the frequency $\omega_1$, while the interacting droplet model predicts 
that for fixed $\omega_2/\omega_1$, $S_2$ will be a decreasing function of 
$\omega_1$\cite{Weissman92,Weissman93}. A sketch of this is shown
in Figure \ref{fig:S2}.
Measurements of the second spectrum of resistance fluctuations in 
the spin glass CuMn find that its behavior is consistent with the 
hierarchical model\cite{Weissman92,Weissman93}. Measurements have
also been done of the second spectrum of the resistance noise in
silicon MOSFETs near the metal-insulator 
transition\cite{Jaroszynski02}. The electron system freezes into a
glassy phase that is consistent with the hierarchical picture
according to the second spectrum. 

It may be possible to use a similar approach with interacting 
two level systems that are in a glassy phase. By looking at
the second spectrum of fluctuations in the dielectric function,
it may be possible to distinguish between the interacting droplet
model and the hierarchical model as ways of describing the
glassy state. 

Experimental measurements of conductance 
fluctuations in the temperature range from 4--30 K
in nanometer--scale samples of amorphous conductors 
(C-Cu and Si-Au) find second spectra that are non-white, indicating
interactions between the fluctuators \cite{Garfunkel89,Garfunkel90}.
Low frequency modulations of the higher frequency noise power
indicate interactions between slow and fast fluctuators.
The density of fluctuators ($1\times 10^{17}/$cm$^{3}K$)
is consistent with the concentration of 
two level systems ($3\times 10^{17}/$cm$^{3}$ K)
estimated from the heat capacity of amorphous 
insulators \cite{revphillips} and superconducting amorphous
metals \cite{Grondey83}. 

A different approach to noise and the second spectrum
due to interacting two level systems
has been given by Nguyen and Girvin\cite{Nguyen01}.
They consider the dynamics of a model with infinite range spin--spin 
interactions. In addition, in their model the tunneling barrier 
height of a two
level system is modulated by interactions with other fluctuating 
two level systems. There are both correlated fluctuations as well as
uncorrelated fluctuations in the barrier heights at different sites.
The correlated fluctuations cause all the barriers to collectively
and simultaneously increase or decrease in a correlated manner. They
speculate that such correlated fluctuations could be produced by the 
elastic interactions between two level systems. In the thermodynamic
limit only the correlated fluctuations survive the central limit
theorem and produce a non--white, non--Gaussian second spectrum.
The other terms produce only Gaussian fluctuations corresponding to
a frequency independent second spectrum. 
%In a more realistic
%system elastic interactions fall off as $1/r^{3}$ and therefore
%fluctuations between two level systems would not be correlated enough
%to survive the thermodynamic limit.

\section{JOSEPHSON JUNCTION QUBITS}
Noise due to fluctuating two level systems has attracted the attention
of those working to make quantum computing qubits, especially
those researchers who are using Josephson junctions in their
qubit design. The Josephson junction qubit is a leading 
candidate in the design of a quantum computer, 
with several experiments recently demonstrating single qubit preparation,
manipulation, and measurement\cite{Vion02,Yu02,Martinis02,Chiorescu03},
as well as the coupling of qubits to each other\cite{Pashkin03,Berkeley03}. 
A significant advantage of this approach is scalability,
as these qubits may be readily fabricated in large numbers using
integrated-circuit technology. A major obstacle to the realization of
quantum computers with Josephson junction qubits is decoherence.

There are a number of ways to design a qubit involving Josephson junctions.
Perhaps the simplest one is a phase qubit\cite{Martinis02} which is
a current biased Josephson junction.
The $|0\rangle$ and $|1\rangle$ states
of the qubit are simply the lowest 2 states in one of the potential
wells of the washboard potential associated with the current biased
Josephson junction. Another design is that of the flux
qubit\cite{Friedman00}. The simplest model of
a flux qubit is an rf SQUID which has one Josephson junction.
In this case the $|0\rangle$ and $|1\rangle$ states
of the qubit are simply the lowest 2 states in the shallower potential
wells of the double well potential of the flux biased SQUID.
The third type of JJ qubit is a charge qubit in which a tiny
superconducting island, the "Cooper pair box", is coupled to a
superconducting reservoir via a Josephson junction. The qubit states
correspond to charge states in the Cooper pair box that differ by
one electron pair\cite{Nakamura99}.
For any of these designs the wavefunction $\psi$ of the qubit is a coherent
superposition of these two states:
\begin{equation}
\psi= \cos\frac{\theta}{2}|0\rangle + \sin\frac{\theta}{2}e^{i\phi}|1\rangle
\end{equation}
Anything that leads to decoherence of this superposition is an
anathema to operation of the qubit.

One source of decoherence are two level systems sitting in
the insulating oxide tunnel barrier of the Josephson junction. 
Fluctuations in a two level system produces fluctuations in the
barrier height which in turn produce fluctuations in the tunneling
matrix element through the oxide barrier and the critical 
current $I_o$\cite{Martinis04}. The 
energy splitting of a phase qubit depends on the size of the critical
current, and so noise in the critical current leads to fluctuations
in the energy splitting of the qubit as well to dephasing. Since
fluctuations in a two level system can lead to fluctuations in the
qubit energy splitting, the two level system and qubit are coupled.
This coupled system has four energy levels. If the qubit and two
level system have the same energy splittings when they are
uncoupled, then the coupling between the qubit and two level 
system will split the degeneracy. Such an energy splitting has
been observed experimentally\cite{Martinis04}.

\section{CONCLUSIONS}
The point of this paper is to urge experimentalists to measure the noise
in their measurements of the low temperature properties of glasses.
Noise coming from fluctuating two level systems can further our
understanding of two level systems in a number of ways. These
include determining if the fluctuations are correlated or 
independent, determining if the two level systems undergo a 
glass transition or a phase transition
with decreasing temperature, and if so, what model best describes
the glassy state, and determining the lower bound of the
two level system relaxation rate (or $\Delta_{o}^{min}$).
In addition understanding the noise produced by two level systems
is important for technological applications such as Josephson
junction devices and qubits that are made from them. 

Most experimental measurements of noise involve electrical measurements
of resistance or current or voltage. Measurements of conductance
fluctuations in amorphous metals indicate that two level systems
interact \cite{Garfunkel89,Garfunkel90}. For insulating glasses
one would measure noise in the dielectric polarization.
Measurements of the noise in the dielectric function are quite challenging
experimentally, but I think they would be well worth the effort.
Measurements of dielectric fluctuations have been done near the
molecular glass transition of polylvinylacetate using atomic
force microscopy techniques that are sensitive to the local
dielectric polarization of the sample\cite{Russell00}.
 
Finally let me wish Professor Hunklinger ``A Very Happy 65th Birthday!''.

\section*{ACKNOWLEDGMENTS}
This work was supported by DOE grant DE-FG02-04ER46107 and 
ONR grant N00014-04-1-0061.

%\bibliography{hunklinger}

\end{document}